\documentclass[twocolumn,showpacs,preprintnumbers,amsmath,amssymb,superscriptaddress,prl,english]{revtex4-2}
\usepackage{epsfig,amsopn}
\usepackage{amssymb,epsfig}
\usepackage{epsfig}
\usepackage{graphicx}
\usepackage{cancel}
\usepackage{ulem}
\graphicspath{{figure/}}
\usepackage{color,xcolor}
\usepackage[caption=false]{subfig}
\usepackage{amsmath,amssymb}
\usepackage{amsthm}
\usepackage{enumerate}
\usepackage[T1]{fontenc}
\usepackage[utf8]{inputenc}
\usepackage{mathtools}   
\usepackage{empheq}
\usepackage{float}

\newcommand{\comment}[1]{}
\newcommand\beq{\begin{equation}}
\newcommand\eeq{\end{equation}}

\newcommand\pa{\parallel}
\newcommand\pr{\perp}

\begin{document}

\title{Chiral detection of Majorana bound states at the edge of a  quantum spin Hall insulator}

\author{Vivekananda Adak}
\affiliation{Department of Physical Sciences, IISER Kolkata, Mohanpur, West Bengal 741246, India}
\author{Aabir Mukhopadhyay}
\affiliation{Department of Physical Sciences, IISER Kolkata, Mohanpur, West Bengal 741246, India}
\author{Suman Jyoti De}
\affiliation{Harish-Chandra Research Institute, A CI of Homi Bhabha National Institute, Chhatnag  Road, Jhunsi, Prayagraj 211019, India}
\author {Udit Khanna} 
\affiliation{Department of Physics, Bar-Ilan University, Ramat Gan 52900, Israel}
\author{Sumathi Rao}
\affiliation{Harish-Chandra Research Institute, A CI of Homi Bhabha National Institute, Chhatnag  Road, Jhunsi, Prayagraj 211019, India}
\affiliation{International Centre for Theoretical Sciences (ICTS—TIFR), Shivakote, Hesaraghatta Hobli, Bangalore 560089, India}
\author{Sourin Das}
\affiliation{Department of Physical Sciences, IISER Kolkata, Mohanpur, West Bengal 741246, India}

\begin{abstract}
A hybrid setup consisting of a superconductivity-proximitized quantum spin Hall (QSH) insulator  and a quantum anomalous Hall (QAH) insulator is proposed for chiral injection of electrons into the Majorana bound state (MBS). An unexplored region of the phase space involving the Zeeman-field induced boost of  the helical edge state is then proposed for the detection of the MBS. 2-D transport simulations of our proposed setup are compared with the corresponding setup in the absence of  the QAH region, when moderate potential and magnetic disorder are included. The remarkable contrast between the two results demonstrates the possibility for an unprecedented immunity from disorder-induced masking of the MBS detection in our proposed setup.
\end{abstract}

\maketitle
{{\underline{\it Introduction}}\,:}
The unambiguous detection of Majorana bound state (MBS) in a quantum transport measurement has remained a challenge,
ever since the first few attempts made in 2012~\cite{Mourik2012,Rokhinson2012,Deng2012,Das2012}.
The initial experimental attempts were based on the nanowire setup~\cite{Sau2010,Oreg2010} but 
since then the field has evolved and two-dimensional platforms based on helical edge states (HES)~\cite{Wu2006,Xu2006, Maciejko2009} 
of quantum spin Hall (QSH)~\cite{Kane2005-1,Kane2005-2,Bernevig2006,Konig2007,Liu2008,Roth2009,PRL2012} state have 
emerged as an alternative~\cite{FuKane2008, Fu2016, Beenakker2013, Wiedenmann2016, Yacoby2014, Trauzettel2011, Trauzettel2012, Trauzettel2014, Trauzettel2017, Trauzettel2018a, Yazdani2019, Trauzettel2020}. These  platforms for detecting the MBS can be used to implement both 
the litmus tests prescribed for the detection of  the MBS, namely, the $2 e^2/h$ conductance peak~\cite{Mi2013} and the
$4\,\pi$ Josephson effect~\cite{FuKane2009,Yacoby2014}. 

The aim of this work is two fold - (a) to come up with a proposal for the detection of the MBS in the 
helical edge state based on the $2 e^2/h$ conductance peak such that it is immune to disorder 
which could mask its clear signature, and (b) to identify a 2-D parameter space for the detection of  the MBS 
comprising of both the in-plane and out-of-plane components of a Zeeman field which acts 
on the edge of the QSH state  hosting the 1-D topological superconductor induced by the 
proximity to an $s$-wave superconductor.
In the presence of a Zeeman field perpendicular to the spin quantization axis (taken to be in the out-of-plane direction~\cite{Weithofer2013})
a topological transition, from the superconducting to the insulating state, takes places as a 
function of the field strength~\cite{Alicea2012}. In contrast, a Zeeman field parallel to the spin 
quantization axis leads to a topological transition into a gapless superconducting phase once the field 
strength is of the order of the superconducting gap~\cite{Meyer2015}. The combined effect of the two 
fields is captured by a zero bias conductance scan which effectively acts as a phase diagram. 

There exist many formidable challenges in the detection of the MBS formed in the edge states of the QSH via the 
$2 e^2/h$ conductance peak. These can be related to the backscattering of 
electrons induced by a variety of perturbations which include the presence of random 
\begin{figure}[H]
	\centering
	\includegraphics[width=1.1\columnwidth]{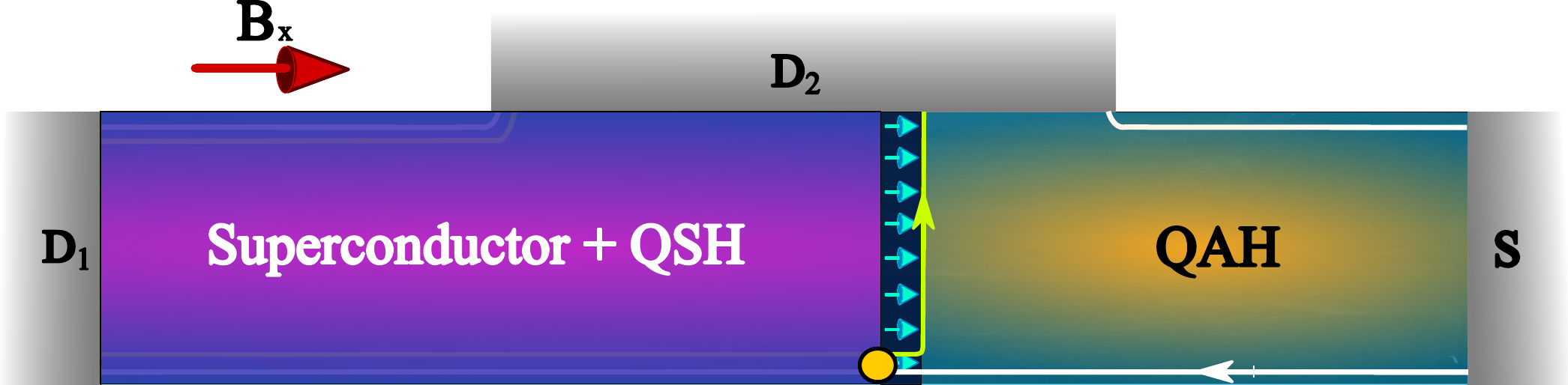}
	\includegraphics[width=0.65\columnwidth]{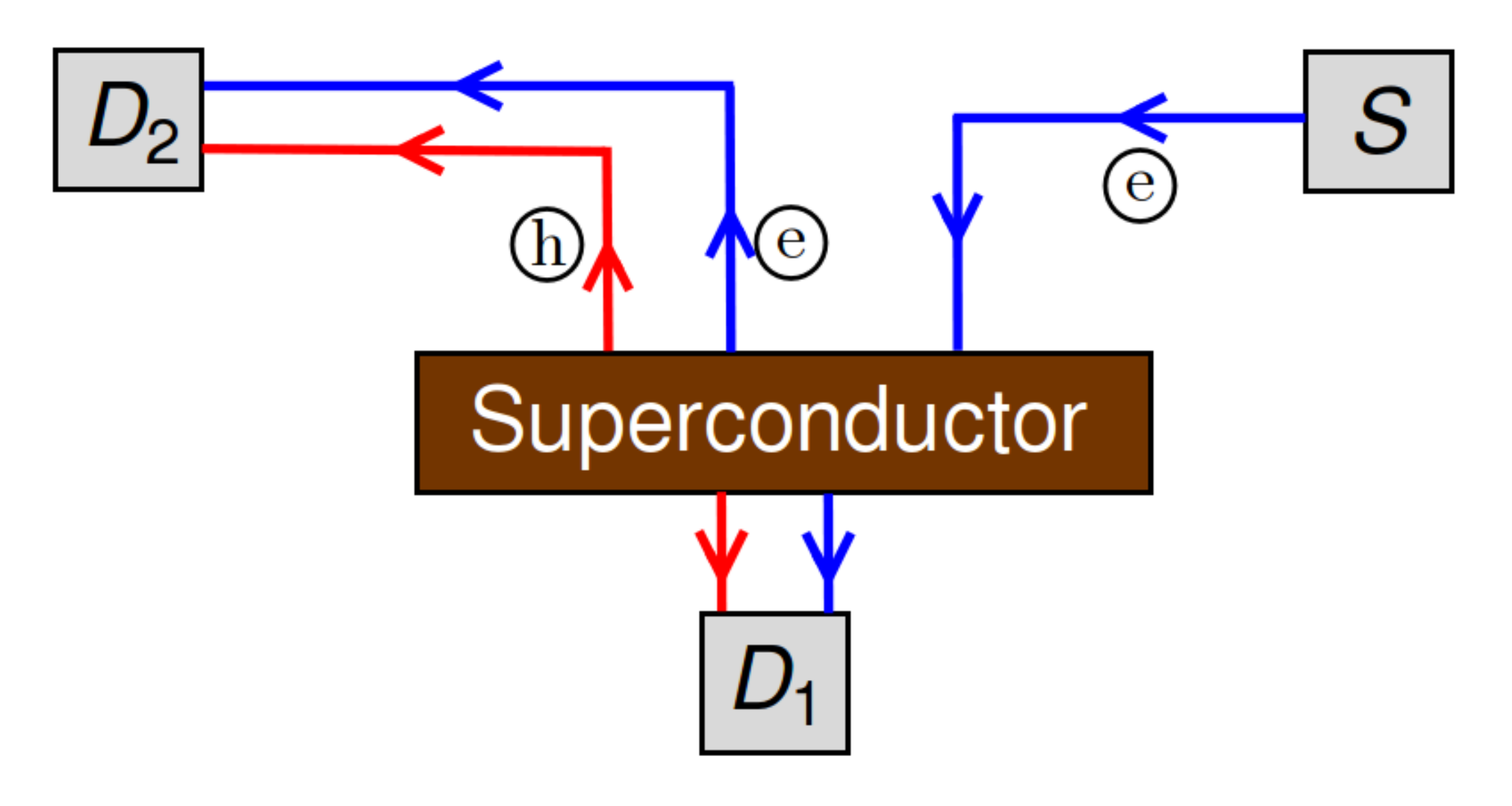}
  \caption{{\it \underline{Top}:} Schematic of the proposed device consisting of a junction between a QSH insulator (purple color) and 
  a QAH insulator (yellow color). The dark region in between (marked with sky-blue arrows) denotes a ferromagnetic barrier. 
  Superconductivity is induced throughout the QSH region through proximity to an $s$-wave grounded superconductor. In the topological 
  phase, a MBS (indicated by the yellow circle) is localized at the boundary between the superconductor and the ferromagnetic barrier (Ferro). 
  The source ($S$) and drains-1,2 ($D_{1,2}$) are three side-coupled contacts employed for the transport measurements. 
  The white and green lines represent the edge modes with opposite spin polarizations. \\
		{\it \underline{Bottom}:} Effective circuit of the setup. The electrons injected by the source (S) are {\it reflected} as electrons and holes at the N-S junction and collected at lead $D_2$. In the gapless phase, electrons can be transmitted through the superconductor and  collected at lead $D_1$. This physical separation of the incoming and reflected channels at the N-S junction is an inherent advantage of the QSH-QAH based setup. 
	}
	\label{figsetup}
\end{figure}
\noindent
exchange fields, magnetic impurities and the Kondo effect, electron-phonon scattering, multi-electron 
scattering due to electron-electron interactions, scattering induced by the coupling to nuclear spins, etc~\cite{Maciejko2009,Tanaka2011,Maciejko2012,Lezmy2012,Budich2012,Schmidt2012, Lunde2012,Eriksson2012,Jukka2013,Del2013,Eriksson2013-1,Eriksson2013-2, Altshuler2013,Geissler2014,Kainaris2014,Jukka2014,Pikulin2014,Dolcetto2016, Kimme2016, Daniel2017, Daniel2018}. 
Such concerns led to theoretical studies  which attempted to spatially separate the left and right movers in the helical edge, and 
hence prevent  backscattering~\cite{Li2013}. This spatial separation in Ref.~\cite{Li2013} was achieved by applying the (out-of-plane) Zeeman
 field on a narrow strip near the edges, which split  the helical edge  into two chiral edges localized at the  two sides of the 
strip by inducing a quantum anomalous Hall (QAH) state between them~\cite{Yang2011,Liu2015}. These studies motivated us to consider a 
setup (shown in Fig.~\ref{figsetup}) which comprises of a junction between a QSH state and a QAH state which facilitates the chiral 
injection of an electron into the MBS, hence providing immunity against backscattering.

{{\underline {\it Phase diagram  in the $g_{\pr}$-\,$g_{\pa}$ plane}}\,:} 
We begin our analysis with the study of topological superconductivity in helical edge states in the presence of Zeeman fields. 
The Hamiltonian of the system is,
\begin{align}
  H_{\text{eff}} = (v_F p_x \sigma_z - \mu) \eta_z + \Delta \eta_x + g_{\pr} \sigma_x + g_{\pa} \sigma_z
\label{H}
\end{align}
where $\sigma (\eta)$ refers to the spin (particle-hole) sector, $\Delta$ is the pairing potential induced 
via  proximity to an $s$-wave superconductor and $g_{\pr}/g_{\pa}$ are the Zeeman terms. 
In the absence of $\Delta$ and $g_{\pr}/g_{\pa}$, $H_{\text{eff}}$ describes a 
single pair of $\sigma_z$ conserving gapless helical channels, which can be thought of as a 1-D massless Dirac
fermion. We assume that the QSH state lies in the x-y plane and hence the spin quantization axis of the 
HES is taken to be along the $z$ direction~\cite{brune2012spin, Nowack2013}.

In the absence of superconductivity ($\Delta = 0$), $g_{\pr}$ hybridizes the two counter-propagating modes and opens up a  mass gap in the spectrum. 
On the other hand, $g_{\pa}$ increases (decreases) the
energy of the right moving spin-up (left moving spin-down) mode without mixing them. This results in a shift of the 
Dirac node from $k = 0$ to $k = -g_{\pa}/v_{F}$, which may be interpreted as providing a {\it boost} to the massless Dirac fermion. 
Physically, the boosted Fermi sea filled up to the Dirac node, supports a persistent charge current due to the finite momentum of the ground state. 
Henceforth in this letter, we shall use the terms mass and boost for the Zeeman terms $g_{\pr}$ and $g_{\pa}$ respectively.

\begin{figure}[t]
\centering
\includegraphics[width=.9\columnwidth]{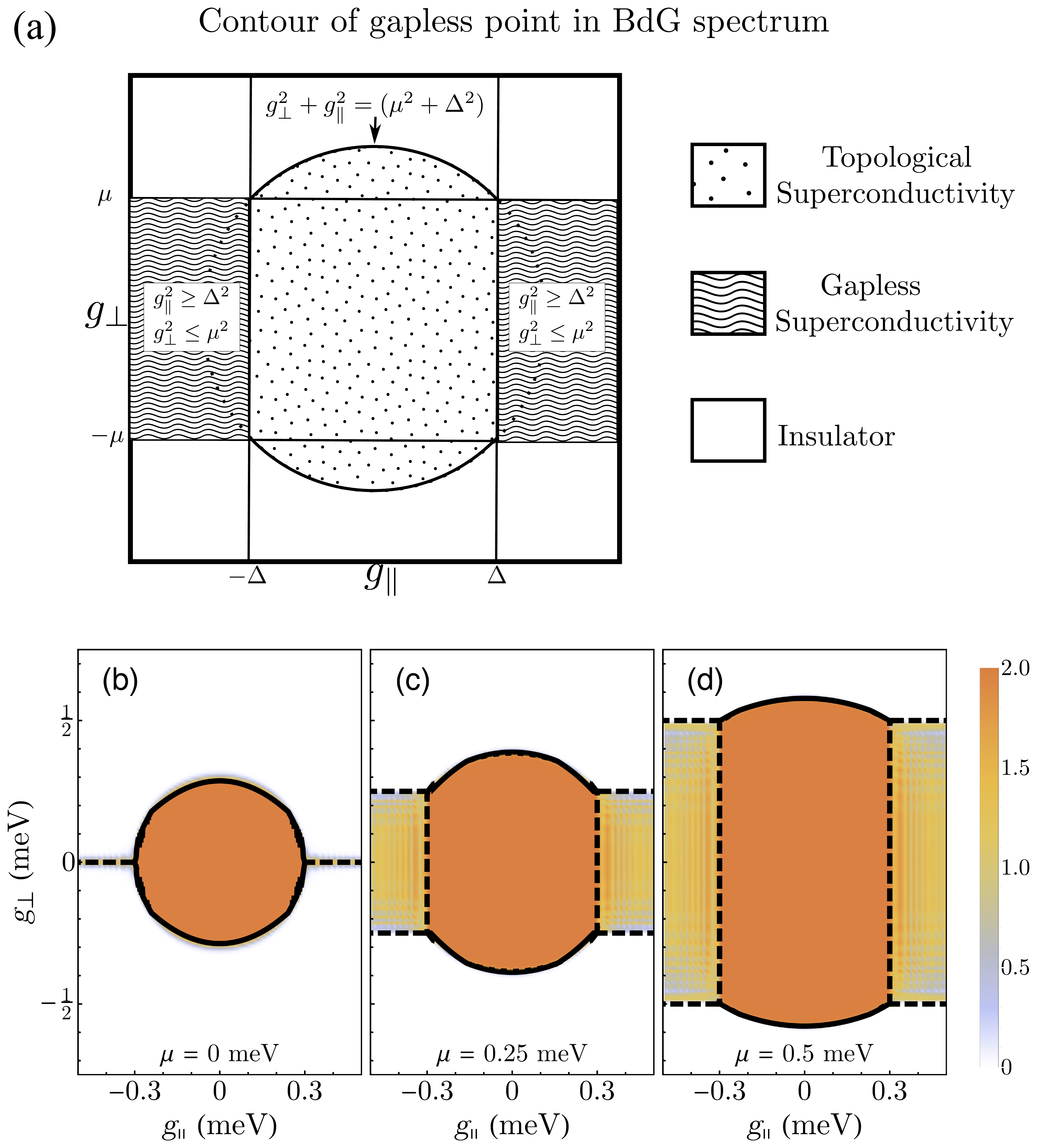}
  \caption{{(a) The phase diagram of HES in the $g_{\pa}$\,-\,$g_{\pr}$ plane. (b-d) Density plots of $1 - R_{ee} + R_{eh}$ (at zero bias) for three different chemical potentials ($\mu = 0.0, 0.25$ and $0.5$ meV). The transport coefficient clearly shows three distinct regions corresponding to topological superconductivity with conductance of $2$ in units of  $e^2/h$, gapless superconductivity with reduced and oscillating conductance,  and an insulating region with zero conductance. The black solid and dashed lines mark the phase boundaries depicted in (a). 
  Clearly the phase diagram based on the BdG spectrum closely matches the {\it transport} phase diagram. } }
\label{figtwo}
\end{figure}

To demarcate the different phases of $H_{\text{eff}}$, we first identify the boundaries separating these phases by solving for the gap closing 
point of the Boguliobov-de Gennes (BdG) spectrum in the parameter space of $g_{\pa}$ and $g_{\pr}$. 
This is given by $\text{Det}\, \lbrack H_{\text{eff}}\rbrack=0$ which reads,
\begin{align}
\label{eq:D}
  k^4 - 2k^2 \lbrack g_{\pa}^2 -g_{\pr}^2 &+ \mu^2 -\Delta^2 \rbrack + \\ \nonumber 
  &\lbrack(g_{\pa}^2 + g_{\pr}^2)- (\mu^2 + \Delta^2)\rbrack^2 = 0 .
\end{align}
From (\ref{eq:D}), we note that $k = 0$ implies $\lbrack(g_{\pa}^2 + g_{\pr}^2)-(\mu^2 + \Delta^2)\rbrack=0$, i.e., 
the circle $g_{\pa}^2 + g_{\pr}^2=\mu^2 + \Delta^2$ defines the locus of the {\it{direct}} gap closing point in the 
parameter space of $g_{\pr}$ and $g_{\pa}$ for given values of $\mu$ and $\Delta$. 
If $\lbrack(g_{\pa}^2 + g_{\pr}^2)-(\mu^2 + \Delta^2)\rbrack \neq 0$, then we have $k= \pm \lbrack (\sqrt{g_{\pa}^2-\Delta^2}) \pm (\sqrt{\mu^2-g_{\pr}^2}) \rbrack$ when $g_{\pa}^2 \geq \Delta^2$ and $ g_{\pr}^2 \leq \mu^2$. This defines the condition of the {\it{indirect}} gap closing, i.e., gap closing at non-zero momentum. The intersection of the region inside the circle $g_{\pa}^2 + g_{\pr}^2=\mu^2 + \Delta^2$ and the area defined by $g_{\pa}^2 \leq \Delta^2$ corresponds to the topological phase hosting a pair of MBS, while the rest of the parameter space either corresponds to gapless superconductivity or an insulating phase [see Fig.~\ref{figtwo}(a)]. It should be noted that the BdG gap closes linearly in $k$ about $k=0$ at the transition at 
$g_{\pr}=\sqrt{\mu^2 + \Delta^2}, \, g_{\pa}=0\,$~\cite{Alicea2012}. In contrast, we find that BdG gap closes quadratically in $k$ about the indirect 
gap closing point at the transition at $g_{\pr}=0,\, g_{\pa}=\Delta$ for finite $\mu$. The rest of the phase boundary presented in Fig.~\ref{figtwo} is a 
result of the interplay of the two gap closing mechanisms governed by the mass and the boost. 
We will see later that the  boost driven transition is very different from the mass driven transition, as the former remains sharply defined even 
in presence of disorder in $\mu$, due to the fact that the condition for this transition is insensitive to $\mu$. 
This fact can have importance for the experimental detection of MBS in our proposed setup. 
This phase diagram, in the presence of the boost and its detection via transport is one of the  key points of this letter.

{{\underline {\it Transport signatures of the  phases of the HES}}}\,: 
Now we  characterize the different phases that we discussed above via  the  transport properties of the setup given in Fig.~1. 
The bottom panel of Fig.~1 shows the effective circuit diagram of the setup. Our interest is in the current flowing from  
$S$ to $D_2$, which collects the electrons and holes reflected from the superconductor. Note that $D_1$ also collects current 
in the gapless phase of the superconductor, since unlike a standard superconductor, the incident electron is not completely 
reflected at the junction.
The 2-D QSH-QAH based setup allows a clear separation of the incoming and the reflected channels, which is not possible in 
semiconductor nanowires or {\it{only}}-QSH based devices. 
We start with an effective edge model which will later be substantiated by 2-D transport simulations. 

We start by calculating the conductance of a normal-superconductor-insulator-normal (N-S-I-N)
junction in the HES where the superconducting region is described by Eq.~(1). The insulating region is modelled as a
ferromagnetic barrier, described by Eq.~(1) with the addition of $M_{x} \sigma_x$. We define 
$\Delta (x) = \Delta \Theta(-x) \Theta(4L+x)$ and 
$g_{\pr/\pa} (x) = g_{\pr/\pa} \Theta(-x) \Theta(4L+x)$ and $M(x) = M_x \Theta(x) \Theta(L-x)$.
We evaluate the probabilities  of normal ($\mathcal{R}_{ee}$) and Andreev ($\mathcal{R}_{eh}$) reflections at 
the junction as a function of the energy of the incident electron~\cite{SM}, particularly focussing 
on energies less than $\Delta$. 

Figures~2(b,c,d) show the plots of $1 - \mathcal{R}_{ee} + \mathcal{R}_{eh}$, which is the two terminal Andreev conductance as a 
function of the mass and the  boost at zero bias for different values of $\mu$. In the region of topological superconductivity 
(shown in deep orange), the conductance is quantized to be $2 e^2/h$ due to the exact resonance with the MBS localized at the junction 
(see also,~\cite{RefereeRequest}). 
In the trivial region (shown in white), Andreev reflection is quenched and the incident electron is fully reflected as an electron. 
This is because without the MBS, the trivial superconductor acts as an insulating barrier at zero energy. As expected for the 
topological transition, the change from  $2 e^2/h$ to 0 occurs at $g_{\pr} = \pm \sqrt{\Delta^2 + \mu^2}$ along the $g_{\pr}$ axis, 
while it falls to values less then $2 e^2/h$ on the $g_{\pa}$ axis at $g_{\pa}=\Delta$. 
The conductance oscillates in the region of gapless superconductivity (shown in mustard yellow), due to the finite probability of 
transmission through the superconductor and the finite size of the superconducting region.

\begin{figure}[t]
\centering
\includegraphics[width=0.95\columnwidth]{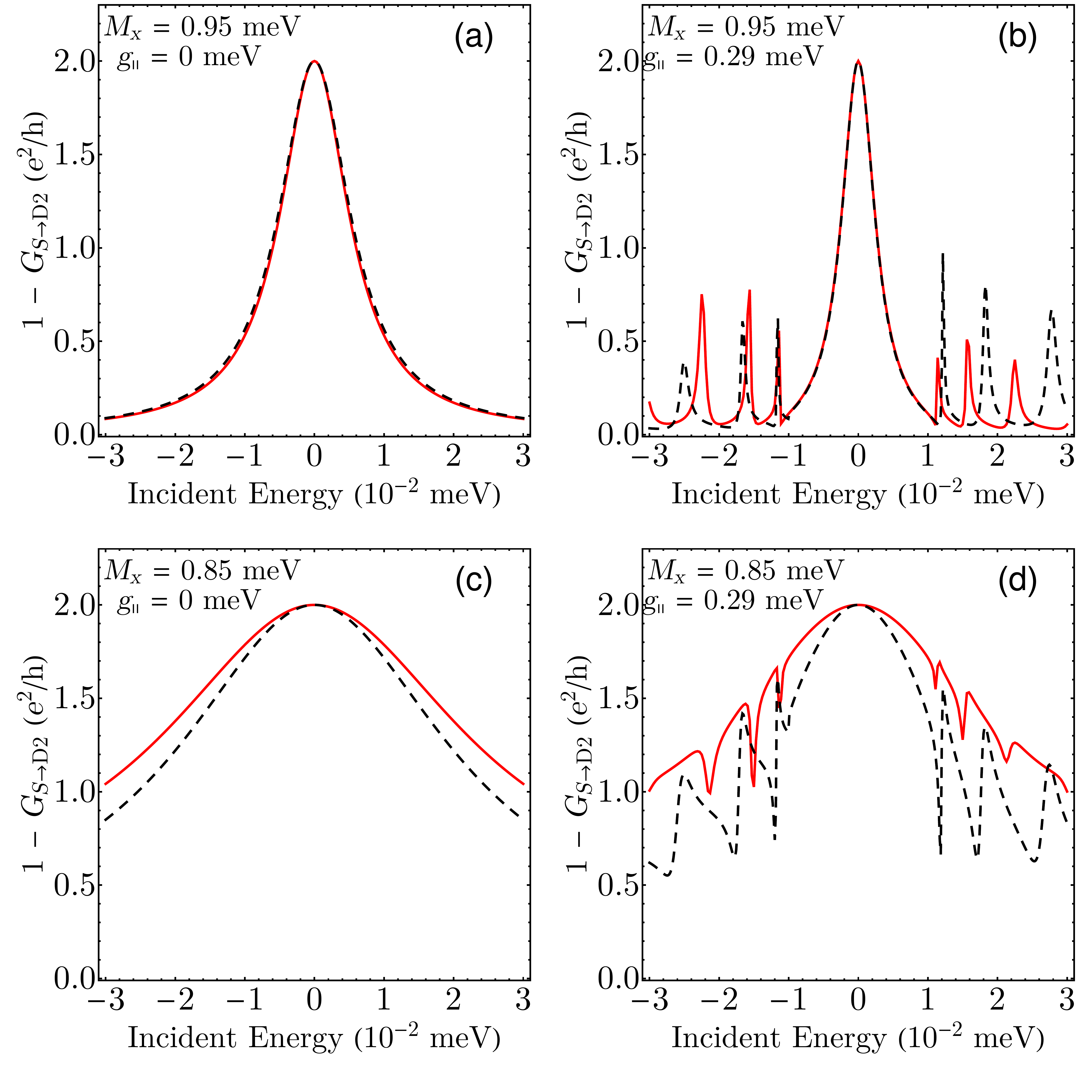}
  \caption{Results from numerical simulations of the setup in Fig.~\ref{figsetup} (in the absence of disorder) for different 
  values of boost ($g_{\pr}$) and amplitude of the ferromagnetic barrier ($M_{x}$). The panels compare the Andreev conductance 
  [$1 - G_{S \rightarrow D2}$ (in units of $e^2/h$)] evaluated from the effective one-dimensional model (the black dashed curves) and
  from the numerical simulation (the red curves). The MBS in the topological regime leads to the zero energy peak.}
\label{figthree}
\end{figure}

{\it{\underline{Numerical Analysis with  the $2$-$D$ model}}\,:}  
The 2-D simulations are performed using the KWANT package~\cite{groth2014kwant} for a device shown in Fig.~\ref{figsetup}. 
The device is modelled by a discretized version of the BHZ model~\cite{Vivek2022}, complemented with the required additional 
terms, over a $700 \times 140$ square lattice 
(with lattice spacing 10 nm). The QAH, superconducting and magnetic regions are defined through the addition of
exchange, pairing potential and Zeeman terms in the appropriate regions~\cite{SM}. We assume that a bulk $s$-wave superconductor is tunnel-coupled 
to the entire QSH region, proximitizing the helical edge modes as well as the bulk. However, as long as the bulk gap of the QSH insulator is 
much larger than the induced pairing amplitude, the superconductor only affects the gapless edge modes. Similarly, the boost and mass terms are 
present in the entire device. The three leads, side-coupled to the device, are also described through the BHZ Hamiltonian. 
For the results presented in Figs.~3-5, we used $\mu = 1.0$ meV, $\Delta = 0.3$ meV and $M_{x} = 0.95$ or $0.85$ meV. The remaining BHZ 
parameter values were close to those describing InAs/GaSb/AlSb quantum wells~\cite{Liu2008,LIU201359,SM}. 

In order to benchmark the device, we first compute the differential conductance from the source to drain $D_2$ at different 
values of the boost, mass and other device parameters. 
Fig.~3 shows $1 - G_{S \rightarrow D2}$ for two values of the boost ($0$ and close to $\Delta$), as well as for two values of the
strength of the ferromagnetic barrier ($M_{x}$). We compare 
the numerical results (shown with red solid curves) to the analytical results (shown with black dashed curves) found from the 
effective edge model. The resonance of $2 e^2/h$ at 
zero energy is evidently the signature of a MBS where its sharpness is dictated by the ferromagnetic barrier. 
As $g_{\pa}$ approaches $\Delta$, there are additional resonances arising from other sub-gap bound states. Note that the theory and 
numerics  match closely but not exactly for the finite energy states. On the other hand 
the results match perfectly for the zero energy resonance. This indicates that the resonance at zero energy is universal while the finite energy 
resonances are sensitive to details (such as dimensionality). The universality is further evidence that the zero energy feature 
is due to a MBS rather than spurious Andreev bound states.

\begin{figure}[t]
\centering
\includegraphics[width=0.99\columnwidth]{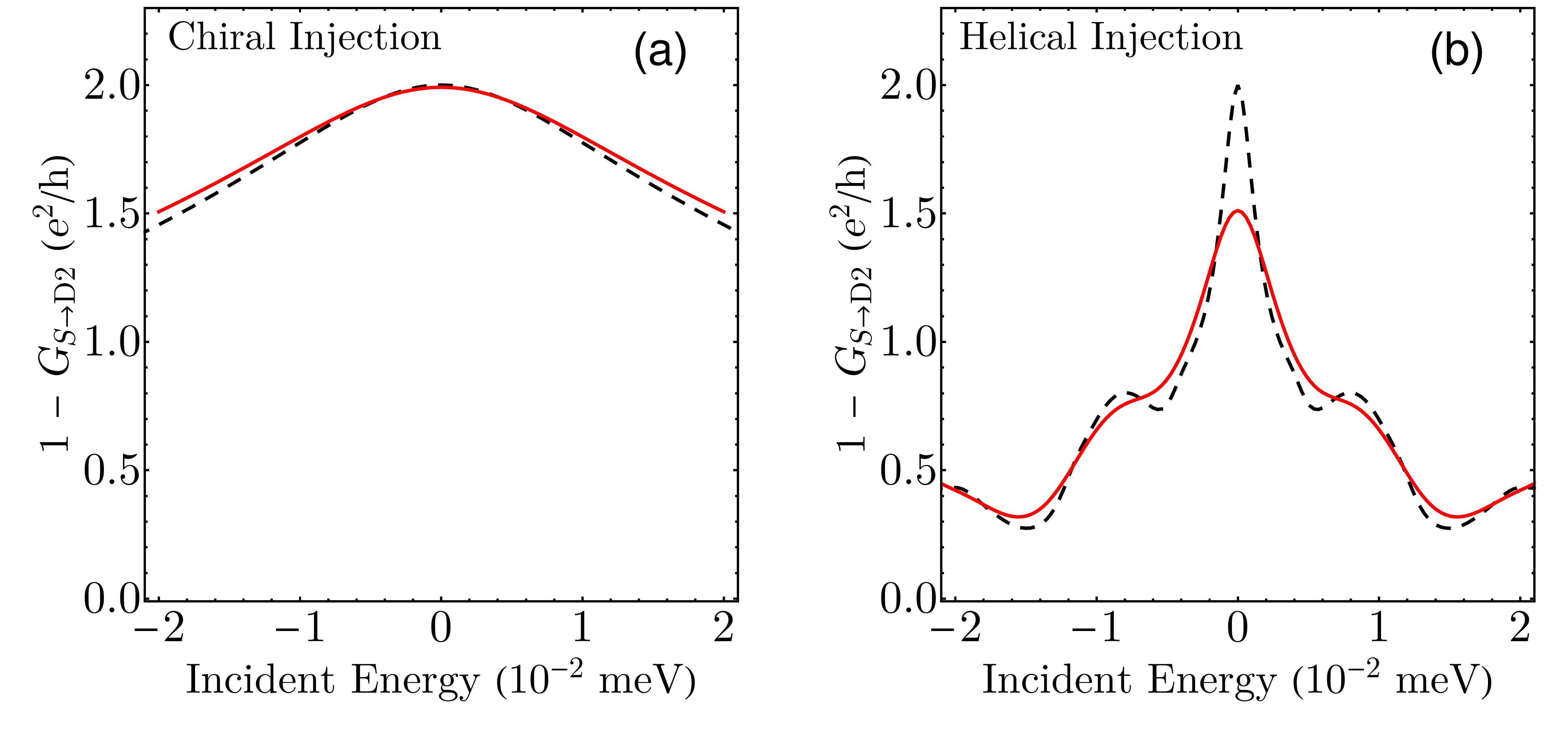}
  \caption{Comparison of the influence of disorder and finite temperature on the Andreev conductance through setups involving injection through a (a) QAH and (b) QSH region. The black dashed [red] curves show (disorder averaged) $1 - G_{S \rightarrow D2}$ (in units of $e^2/h$) in the presence of potential (time-reversal invariant) and ferromagnetic (time-reversal breaking) disorder at $T = 0$ [$T \sim 12$ mK]. Note that at $T = 0$, the zero-energy conductance is quantized to $2 e^{2}/h$ independent of the nature of injection, but the width of the resonance is greatly reduced in case of helical injection through a QSH region due to disorder-induced backscattering. At finite $T$, thermal smearing reduces the peak height and increases the peak width in both cases. However, this smearing is significantly suppressed with chiral injection. Thus, demonstrating the central advantage of our proposal. }
\label{figfour}
\end{figure}


{\it {\underline {Robustness to disorder}}\,:} The analysis presented above is based on a disorder free setup. 
A key advantage of our proposal over designs based on one-dimensional nanowires or helical edges of the QSH phase is that the 
junction with the QAH region allows for chiral injection of electrons. The chiral nature of electrons incident upon the N-I-S 
interface provides strong immunity from adverse effects of disorder induced backscattering, particularly in the presence of thermal 
smearing (at finite temperatures). While the helical edge modes of the QSH phase are expected to be robust to time-reversal 
symmetry preserving weak disorder, time-reversal breaking induced backscattering sharply reduces the width of the zero bias peak 
and makes detection difficult~\cite{Beenakker2013}. We will demonstrate that the resonance height and width are much more robust in the 
case of chiral injection, which considerably increases the feasibilty of detection.

\begin{figure}[t]
\centering
\includegraphics[width=\columnwidth]{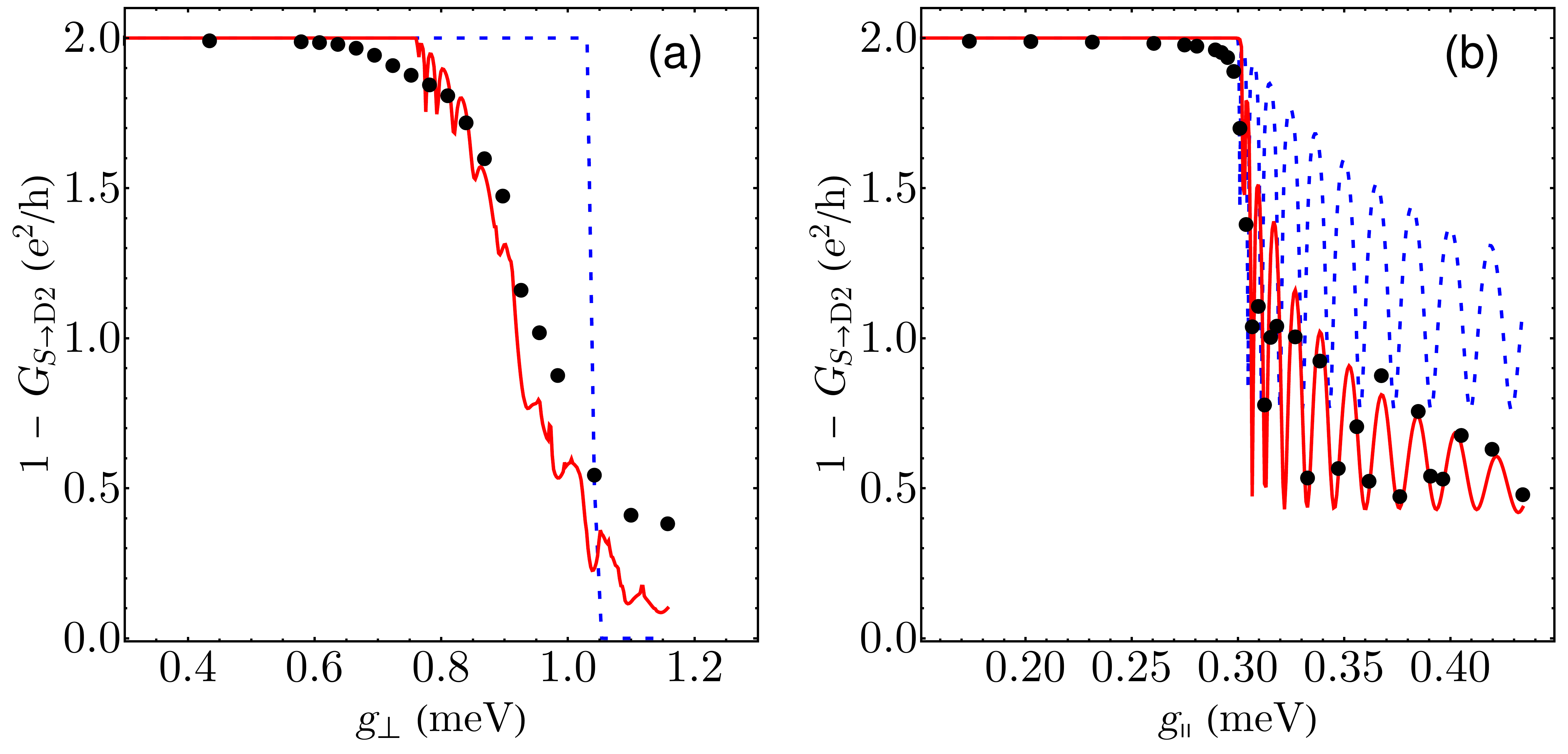}
  \caption{Influence of disorder and finite temperature on the sharpness of the phase transitions at finite (a) mass ($g_{\pr}$) and (b) boost ($g_{\pa}$). The blue dashed curves show the zero energy Andreev conductance ($1 - G_{S1 \rightarrow D2}$) at $T = 0$ in the absence of disorder. Our analytical model predicts phase transitions at $g_{\pr} \sim 1.0$ meV [in panel (a)] and $g_{\pa} \sim 0.3$ meV [in panel (b)]. The red curves (black dots) show the disorder averaged Andreev conductance at $T = 0$ ($T \sim 12$ mK). Clearly, the transition at finite mass [panel (a)] is significantly broadened due to the disorder and thermal effects, while the transition at finite boost [panel (b)] remains fairly robust. }
\label{figfive}
\end{figure}

We repeated the numerical analysis of our setup in presence of both potential (time-reversal invariant) and ferromagnetic 
(time-reversal symmetry breaking) disorder to verify the robustness of our proposal. To model a disordered 
sample, a weak random onsite potential was introduced throughout the setup. 
We also included weak magnetic disorder, in the form of a random onsite spin-nonconserving potential, throughout the setup to 
break the time-reversal symmetry and allow scattering between spin-up and spin-down states.
We performed the numerical simulations for the disordered setup at both zero and finite temperatures. 
Finally, we performed identical simulations for a similar setup, adapted from Ref.~\cite{Beenakker2013}, in which the QAH region of Fig.~1
is replaced by a QSH region (i.e. the source is coupled directly to the helical edge mode of a QSH phase). This was done in order to highlight 
the efficiency of a device with chiral injection (through a QSH-QAH junction) in reducing the effects of disorder and thermal smearing.

Fig.~\ref{figfour} presents the results of the analysis of disordered setups. Panels (a) [(b)] show the (disordered average) 
Andreev conductance through setups based on chiral [helical] injection, i.e. the device includes [does not include] a QAH region. 
For zero temperature (cf. black dashed curves), although the zero-bias conductance is quantized to be $2 e^{2}/h$, the width of the resonance
differs significantly between the two setups. In the case of 
chiral injection (which is immune from backscattering), the width of the peak is determined by the device details 
(primarily by the height and width of the ferromagnetic barrier), while in the case of helical injection, disorder-induced 
backscattering strongly reduces the peak width. At a finite temperature (cf. red solid curves), the peak height (width) is reduced (increased) 
in both cases. However the effect is significantly suppressed for the case of QAH-based setups, clearly demonstrating the advantage
of chiral injection. 

We also studied the effect of disorder (and finite temperature) on the phase transitions at finite mass/boost described in the 
previous section. Fig.~\ref{figfive} shows the results of these simulations. The blue (dashed) curves depict the zero-bias conductance 
as a function of the mass [Panel (a)] and boost [Panel (b)] for a clean setup at $T = 0$. From our previous analysis, we expect the 
phase transitions to occur at $g_{\pr} = \sqrt{\Delta^{2} + \mu^{2}} \sim 1$ meV and $g_{\pa} = \Delta = 0.3$ meV. Clearly, the numerical
results show very sharp transitions at these values. The red curves depict the (disorder-averaged) zero-bias conductance at $T = 0$. 
The transition along finite boost clearly remains sharp, while the transition along finite mass is smeared out. This remains true 
even at finite temperatures (cf. black dots). The robustness of the boost-driven transition to both disorder and thermal effects 
suggests a clear advantage from an experimental point of view.

{\it {\underline {Summary}}\,: } To summarise, we have proposed a device for observing MBS based on chiral injection 
in a  QSH-QAH interface, which is immune to disorder and paves the way for a clear detection of the $2e^2/h$ zero bias peak arising from the MBS. 
We have also seen that the zero energy resonance is more robust to disorder in the presence of a boost than in the presence of 
a mass gap. These are the two key findings of this letter.


\acknowledgments{{\it {\underline {Acknowledgments}}\,:} VA acknowledges support from IISER Kolkata in the form of a subsistence grant. AM would like to thank Ministry of Education, India 
for financial support.  SJD acknowledges support from Infosys grant. U.K. was supported by a fellowship from the Israel Science Foundation (ISF, Grant No. 993/19). 
S.D. would like to acknowledge the MATRICS grant (MTR/ 2019/001 043) from the Science and Engineering Research Board (SERB) for funding. 
We acknowledge the central computing facility (DIRAC supercomputer) and the computational facility at the Department
of Physics (KEPLER) at IISER Kolkata. SD and VA are grateful to Satyabrata Raj for his help with KEPLER cluster.}


\bibliographystyle{apsrev} 
\bibliography{boostmajorana}
\end{document}


\renewcommand{\thefigure}{S\arabic{figure}}
\setcounter{figure}{0}
\renewcommand{\theequation}{S\arabic{equation}}
\setcounter{equation}{0}
\renewcommand\thesection{S\arabic{section}}
\setcounter{section}{0}

\title{Supplemental material for ``Chiral detection of Majorana bound states at the edge of a  quantum spin Hall insulator''}

\author{Vivekananda Adak}
\affiliation{Department of Physical Sciences, IISER Kolkata, Mohanpur, West Bengal 741246, India}
\author{Aabir Mukhopadhyay}
\affiliation{Department of Physical Sciences, IISER Kolkata, Mohanpur, West Bengal 741246, India}
\author{Suman Jyoti De}
\affiliation{Harish-Chandra Research Institute, A CI of Homi Bhabha National Institute, Chhatnag  Road, Jhunsi, Prayagraj 211019, India}
\author {Udit Khanna} 
\affiliation{Department of Physics, Bar-Ilan University, Ramat Gan 52900, Israel}
\author{Sumathi Rao}
\affiliation{Harish-Chandra Research Institute, A CI of Homi Bhabha National Institute, Chhatnag  Road, Jhunsi, Prayagraj 211019, India}
\author{Sourin Das}
\affiliation{Department of Physical Sciences, IISER Kolkata, Mohanpur, West Bengal 741246, India}

\date{\today}

\begin{abstract}
This supplemental material provides additional details regarding the analytical and numerical analysis of the device proposed in this work.
\end{abstract}

\maketitle

\section*{I. Analytical results}
We consider a HES of a QSH state as described in Eq. (1) of the main text.
\begin{equation}
  H_{\text{eff}}= (v_F p_x \sigma_{z}-\mu)\eta_{z} + \Delta \eta_x + g_{\perp} \sigma_{x} + g_{\parallel} \sigma_z + M \sigma_x,
	\label{hamS}
\end{equation}
where, $M(x) = M_x \Theta(x).\Theta(-x+L)$, 
$g_{\parallel/\perp}(x) = g_{\parallel/\perp} \Theta(-x).\Theta(4L+x)$ and $\Delta(x) = \Delta \Theta(-x).\Theta(4L+x)$. 
Here, we employed the basis: $( \psi_{\uparrow},\psi_{\downarrow},\psi_{\downarrow}^{\dagger},-\psi_{\uparrow}^{\dagger} )$. 
We include an insulating ferromagnetic barrier between the superconductor and the lead to localize the Majorana
zero mode at the interface. We use the notation defined in the main text.
An electron, injected with (excitation) energy $\epsilon$ from $x = \infty$, incident upon the ferromagnet at $x = L$ can undergo normal or Andreev
reflection with amplitudes $r_{ee}$ and $r_{he}$ respectively. Then a general wave function in the normal region 
at $x > L$ can be written as,
\begin{align}
	\psi_{N1} = e^{-i k_e x}
	\begin{bmatrix}
		1	\\
		0	\\
		0	\\
		0	
	\end{bmatrix} +
	r_{ee} e^{i k_e x}
	\begin{bmatrix}
		0	\\
		1	\\
		0	\\
		0	
	\end{bmatrix} +
	r_{he} e^{-i k_h x}
	\begin{bmatrix}
		0	\\
		0	\\
		1	\\
		0	
	\end{bmatrix}
\end{align}
Note that each column vector has been normalized such that it carries unit current. 
Here, $k_e=(\mu+\epsilon)/\hbar v_F$ and $k_h=(\mu-\epsilon)/\hbar v_F$.
Since the superconductor is finite, the electron can also tunnel through the superconductor into the normal 
region at $x < -4L$, as an electron or a hole with amplitudes $t_{ee}$ and $t_{he}$ respectively. A generic wave function
in this region is, 
\begin{align}
	\psi_{N4} = t_{ee} e^{-i k_e x}
	\begin{bmatrix}
		1	\\
		0	\\
		0	\\
		0
	\end{bmatrix} +
	t_{he} e^{i k_h x}
	\begin{bmatrix}
		0	\\
		0	\\
		0	\\
		1
	\end{bmatrix}
\end{align}

The wave function in $-4L < x < 0$ and in $0<x<L$ are the superpositions of all possible solutions at a given energy. 
We compute the solutions in these regions numerically. The Hamiltonian in Eq.~\ref{hamS} is numerically diagonalized to find 
all eigenvectors with the same energy. Imposing continuity of wave functions at the boundaries ($x = L, 0, -4L$), the 
transmission and reflection amplitudes can be evaluated by solving a set of linear algebraic equations. 
The reflection probabilities ($\mathcal{R}_{ee} = |r_{ee}|^2$ and $\mathcal{R}_{he} = |r_{he}|^2$) are readily obtained
and the conductance : $1 - \mathcal{R}_{ee} + \mathcal{R}_{he}$, can be directly compared with the results of the 
numerical simulation of the full 2D setup. 
For the results presented in Figs.~2,3 we used parameters identical to those employed for the 2D simulation, 
namely: $\mu = 1.0, \Delta = 0.3, M_x = 0.5$ (all in meVs). The Fermi velocity $v_{F}$ is obtained by fitting the edge mode dispersion
(from the 2D calculation) to linear functions. This procedure gives $\hbar v_{F} = 37.0$ meV nm$^{-1}$. 
The length $L$ is used as a fitting parameter to fine tune the match between numerical and analytical results. 


\begin{figure}[t]
\centering
\includegraphics[width=0.48\columnwidth]{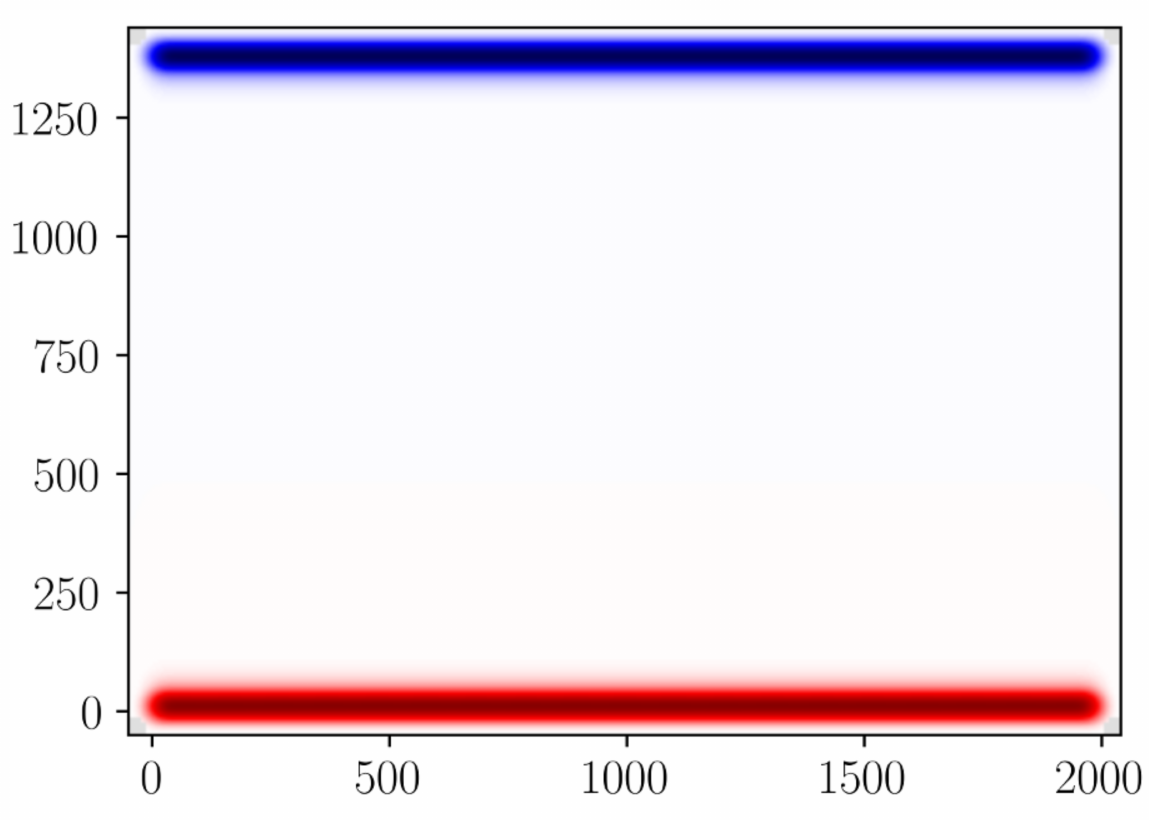}
\includegraphics[width=0.50\columnwidth]{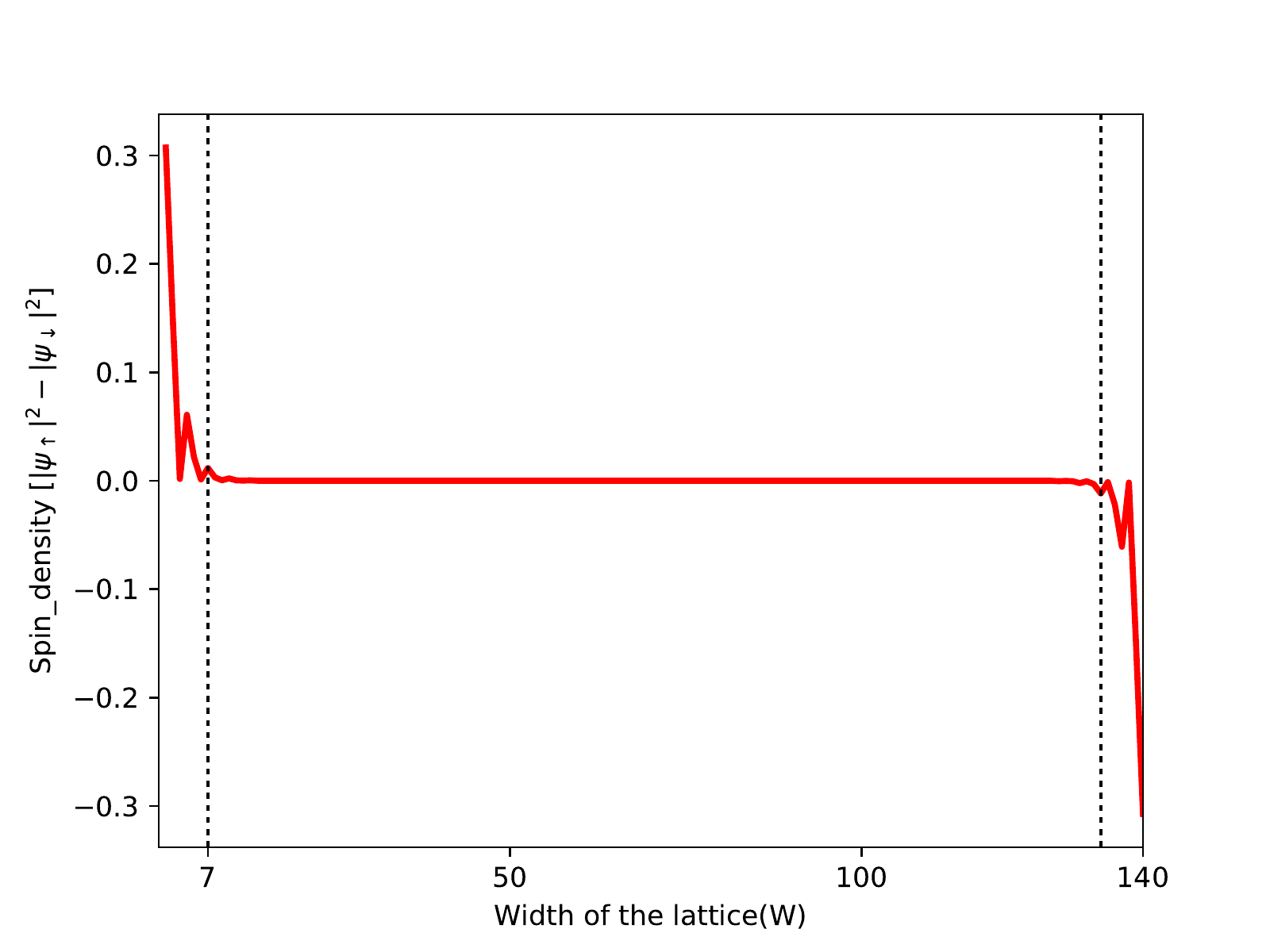}
  \caption{ (Left) Color map of spin density of the QSH phase in a ($210$ nm $\times 420$ nm) 
  two-terminal setup. (Right) Line plot of the spin density at a fixed value of $x$. The opposite 
  spin polarization at the top and bottom edges indicates the presence of the helical edge modes 
  of width $\sim 70$ nm (7 lattice sites). 
  The parameters used here are the  same as those used while simulating the full device.  }
  \label{F-spindensity}
\end{figure}

\section*{II.\,\,\,\,\,\,     Details of the Numerical Simulation}

\subsubsection*{Basic Setup}

We perform a simulation of quantum transport through the two-dimensional (2D) device described in the main text 
[Fig.~1(a)] using the KWANT package~\cite{Kwant}. The system is modelled through a discretized version of the 
BHZ model~\cite{BHZ} complemented with additional terms. The leads are also described using the same Hamiltonian.
In the continuum, the BHZ model Hamiltonian reads, 
\begin{align}
  H_{\text{BHZ}} = \mu -D k^2 + A (k_x \sigma_z \tau_x - k_y \tau_y) + M(k) \tau_z,
\end{align}
where $M(k) = M - B k^2$ and  $\sigma$ ($\tau$) act on the spin (orbital) spaces respectively. $\mu$ is the chemical
potential which can be tuned using an external gate, while $A$, $B$, $D$ and $M$ are material 
dependent parameters. We also include Rashba spin-orbit coupling,
which is naturally present in the two-dimensional quantum wells due to breaking of inversion symmetry. 
Since $H_{\text{BHZ}}$ is block diagonal in spin space, the spin-orbit coupling can be represented by off-diagonal terms,
\begin{align}
  H_{SO} &= 
  \left( \begin{array}{cc}
    0 & H_1(k) \\
    -H_1^*(-k) & 0
  \end{array} \right) \,\, \text{ where,} \\
  H_1(k) &= \left( \begin{array}{cc}
    c_1 k_{+} + i c_3 k_{-} & -c_0\\
    c_0 & c_2 k_{-}
  \end{array} \right).
\end{align}
Here $k_{\pm} = k_{x} \pm i k_{y}$ and the strength of spin-orbit interaction is controlled by 
the parameters $c_{i}$ ($i = 0, 1, 2, 3$). In this work, we use parameter values corresponding to InAs/GaSb/AlSb 
quantum wells~\cite{Models}, namely $A = 37.0$ nm meV, $B = -660$ nm$^2$ meV, $D = 0$ nm$^2$ meV and $M = -7.8$ meV,  
$c_0 = 0.2$ meV, $c_1 = 0.066$ meV nm, $c_2 = 0.06$ meV nm and $c_3 = -7.0$ meV nm~\cite{Models}. The lattice 
constant was assumed to be $a = 10$ nm, and the global chemical potential was set to $\mu = 1.0$ meV. The exchange field 
used to generate the QAH phase was $g_0 = -12.0$ meV, the superconducting order
parameter was $0.3$ meV and the strength of the ferromagnetic barrier was considered as $0.85$ or $0.95$ meV. 
It is important to note that the BHZ model is only an effective model describing the low energy band structure of the quantum well. 
It does not describe the microscopic details of the setup. 
For this reason, the numbers used here for the parameters (such as the lattice constant) may differ from the actual experimental values. 

To study topological superconductivity in the system, we use the BdG representation of 
the BHZ model. Using the basis order $( \psi_{1 \ua}, \psi_{2 \ua}, \psi_{1 \da}, \psi_{2 \da}, \psi^{\dagger}_{1 \da}, 
\psi^{\dagger}_{2 \da}, -\psi^{\dagger}_{1 \ua}, -\psi^{\dagger}_{2 \ua} )$,
[here $1/2$ are the orbitals and $\ua/\da$ are the spin indices along $s_z$ direction] 
the Hamiltonian can be rewritten as
\begin{align}
  H = \mu \eta_{z} -D k^2 \eta_{z} + A (k_x \eta_z \sigma_z \tau_x - k_y \eta_z \tau_y) + M(k) \eta_z \tau_z,
\end{align}
where $\eta$ acts on the particle-hole space. 
The tight-binding version of $H$ is constructed on a square lattice with two 
orbitals per site, through the following substitutions followed by a Fourier transform, 
\begin{align} 
  k^2 &\rightarrow 2 a^{-2}[2-\cos(k_x a)-\cos(k_y a)], \\ 
  k_x &\rightarrow a^{-1} \sin(k_x a), \\ 
  k_y &\rightarrow a^{-1} \sin(k_y a).
\end{align}
In this work, we assume the lattice spacing $a = 10$ nm. Using this real space model, 
we compute the spin density of the ground state at $\mu = 1$ meV (shown in 
Fig.~S1) in a standard two-terminal geometry. The presence of spin polarization at the top and
bottom edges indicates the presence of the helical edge modes of the quantum spin Hall (QSH) phase. 
Fig.~S1 also helps us to find the width of the edge mode for the parameters used here. 

\begin{figure}[t]
\centering
\includegraphics[width=0.49\columnwidth]{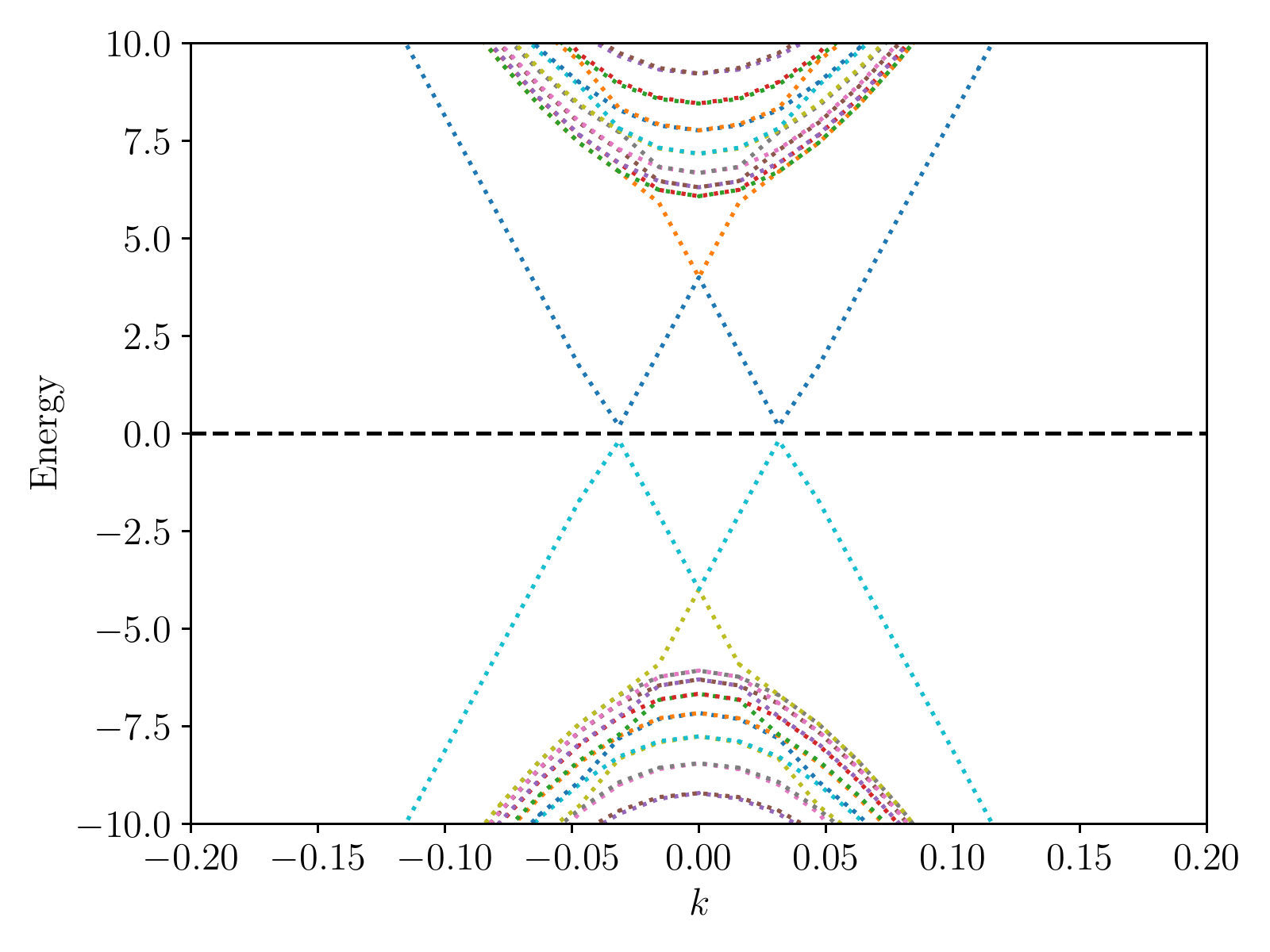}
\includegraphics[width=0.49\columnwidth]{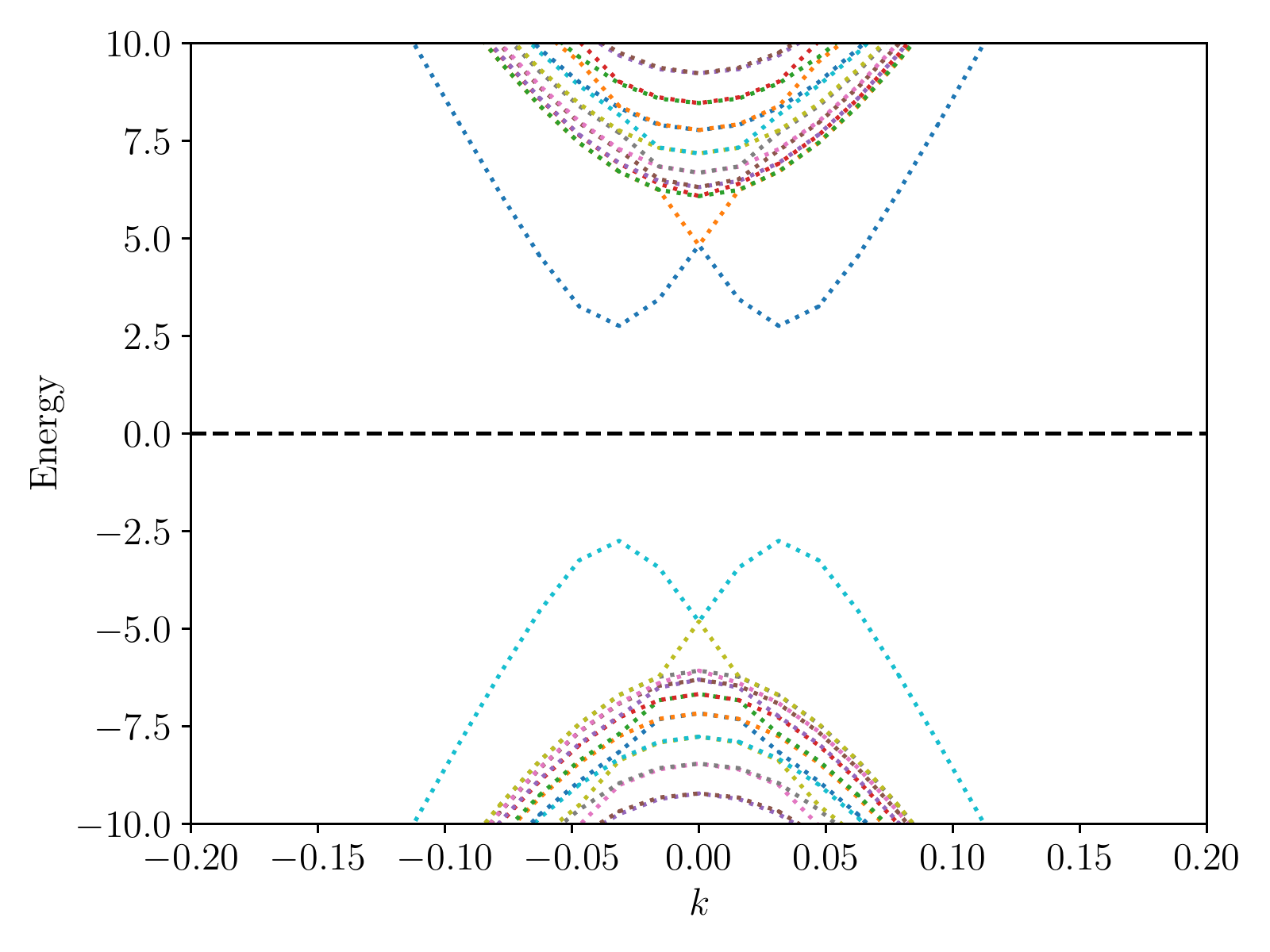}
\includegraphics[width=0.49\columnwidth]{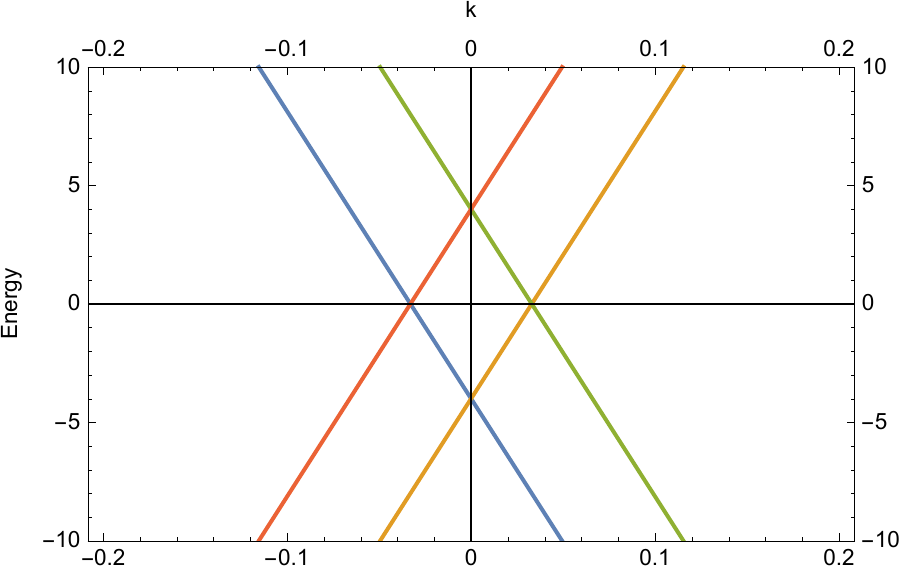}
\includegraphics[width=0.49\columnwidth]{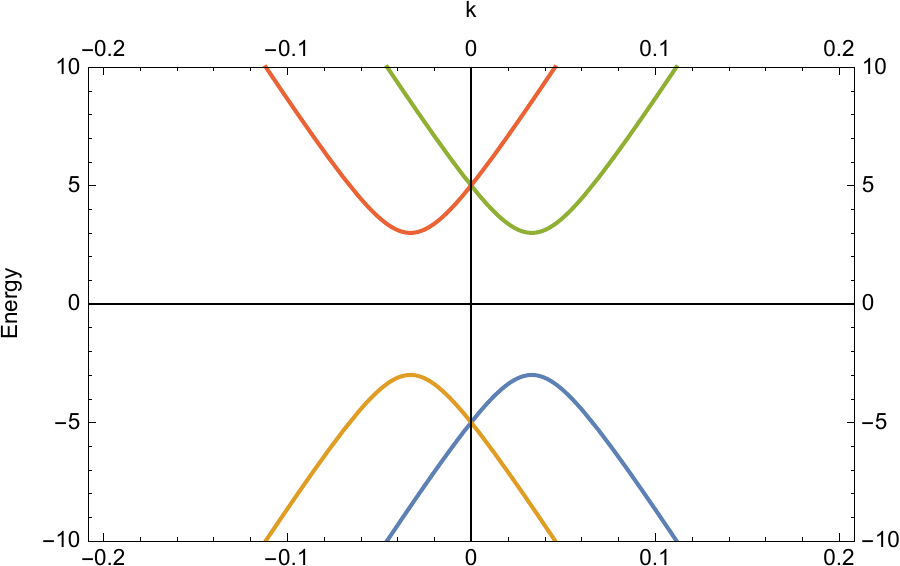}
  \caption{ (Top) Band structure of the 2D model with (left) $\Delta = 0$ and (right) $\Delta \neq 0$ in a 
  semi-infinite cylinderical geometry. (Bottom) Edge mode dispersion found using the effective 1D model
  (described in the main text) with same value of $\Delta$. 
  The energy is in units of meV, while $k$ is in units of nm$^{-1}$. 
  The parameters used here for the 2D model are mentioned in the text. 
  For the 1D model we use $v_F = 37.0$ meV nm$^{-1}$.  }
  \label{F-Bands_Delta}
\end{figure}

\begin{figure}[t]
\centering
\includegraphics[width=0.49\columnwidth]{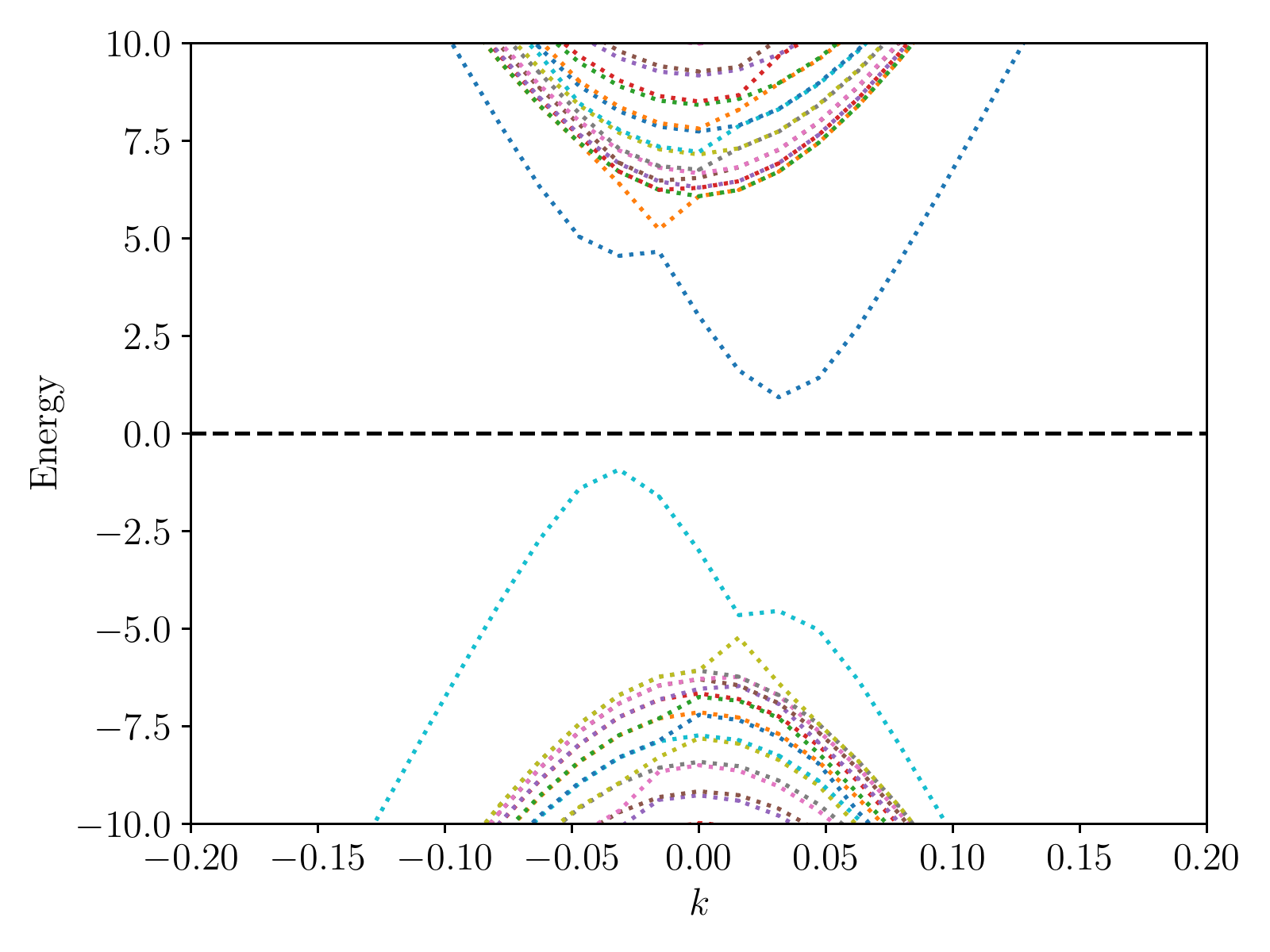}
\includegraphics[width=0.49\columnwidth]{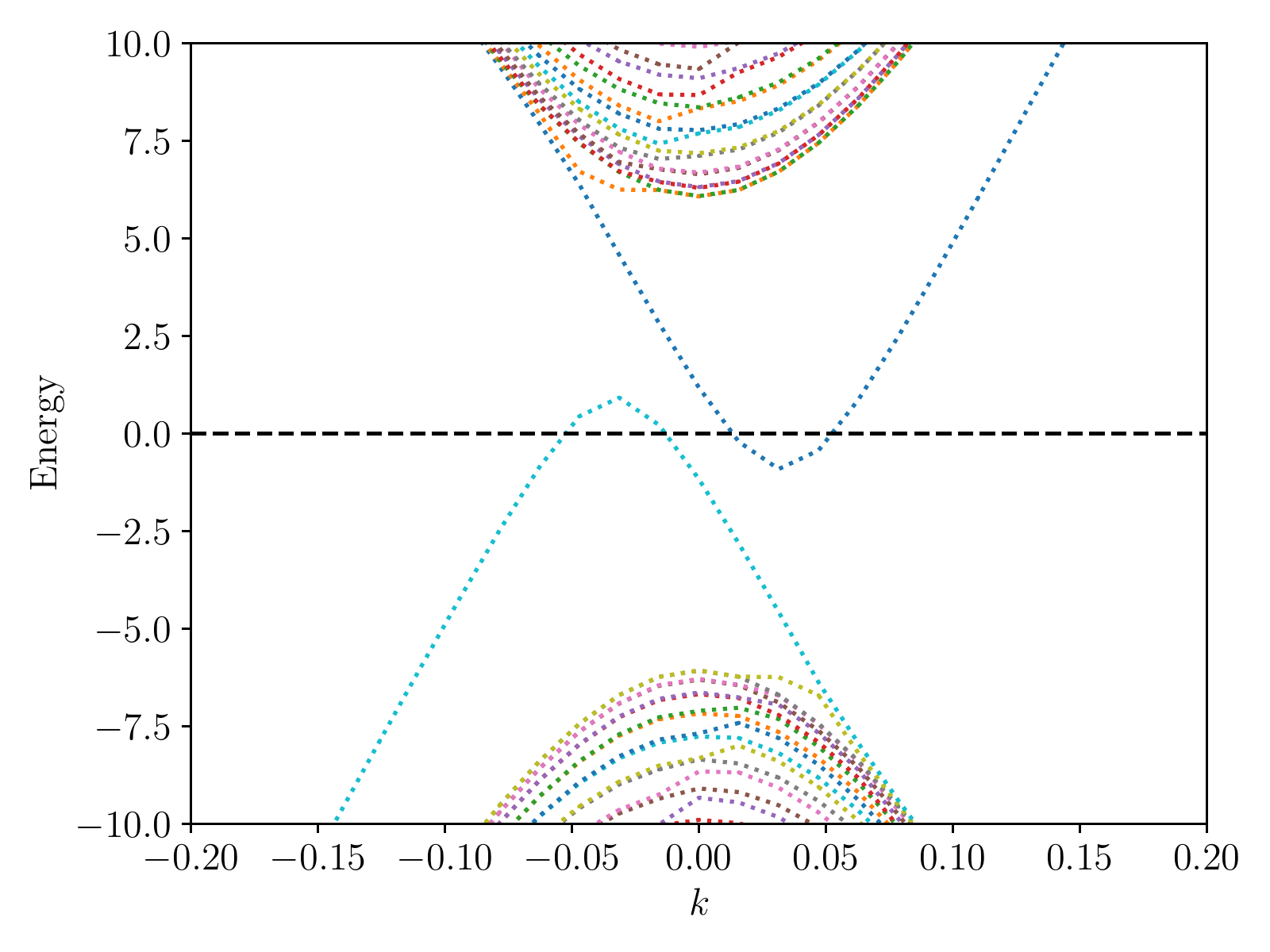}
\includegraphics[width=0.49\columnwidth]{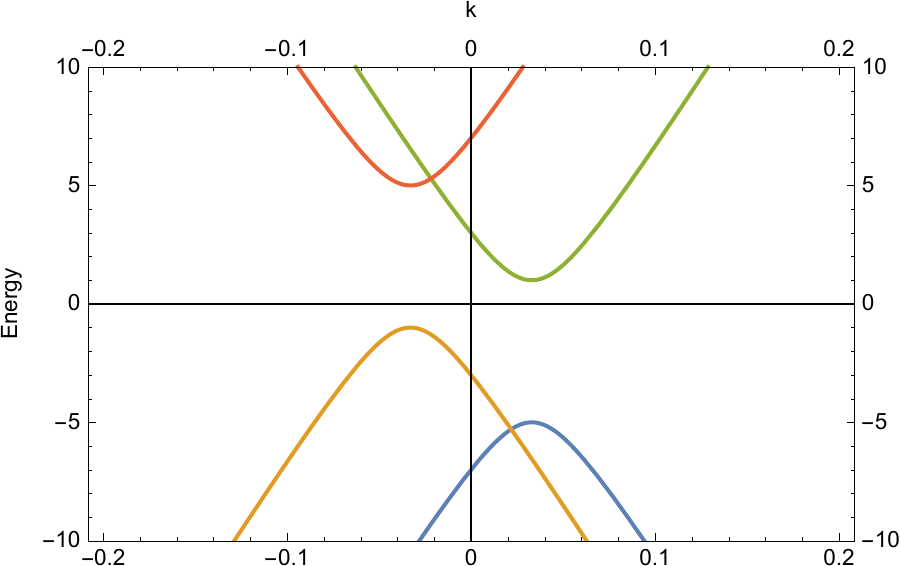}
\includegraphics[width=0.49\columnwidth]{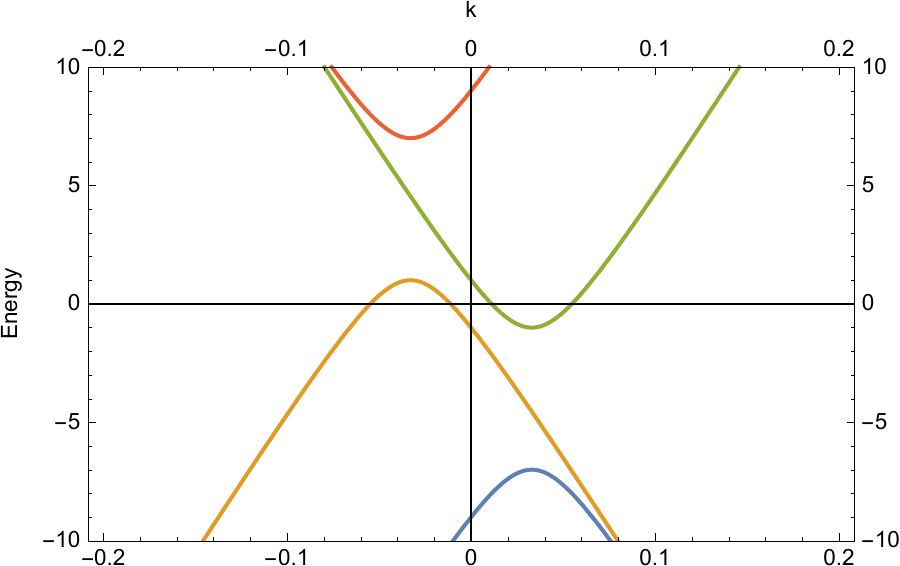}
  \caption{ Evolution of the band structure with boost. (Top) Dispersion of the 2D model with 
  (left) $g_{\parallel} < \Delta$ and (right) $g_{\parallel} > \Delta$ in a semi-infinite cylinderical geometry. 
  (Bottom) Edge mode dispersion found using the effective 1D model
  (described in the main text) with same value of $\Delta$ and $g_{\parallel}$. 
  The energy is in units of meV, while $k$ is in units of nm$^{-1}$. 
  The parameters used here for the 2D model are mentioned in the text.
  For the 1D model we use $v_F = 37.0$ meV nm$^{-1}$.  }
  \label{F-Bands_Boost}
\end{figure}

To model the topological superconductor on the edge, arising from proximity to an $s$-wave superconductor,
we add a pairing potential $\Delta \eta_{x}$ throughout the QSH region. Although the superconductor is tunnel-coupled to both 
the bulk and the edge, for $\Delta \ll M$, only the edge is affected qualitatively. As an additional consistency check, we compute
the band structure of the 2D model in a semi-infinite cylindrical geometry and compare with the 
band structure found using the effective one-dimensional (1D) model described in the main text. Fig.~S2 
shows the bands before [panels (a) and (b)] and after [panels (c) and (d)] adding a finite $\Delta$ 
at the edge. The close match of the subgap (edge) bands of the 2D results with the 1D dispersion 
indicates that the pairing potential does not disturb the bulk bands.

\begin{figure}[t]
\centering
\includegraphics[width=0.90\columnwidth]{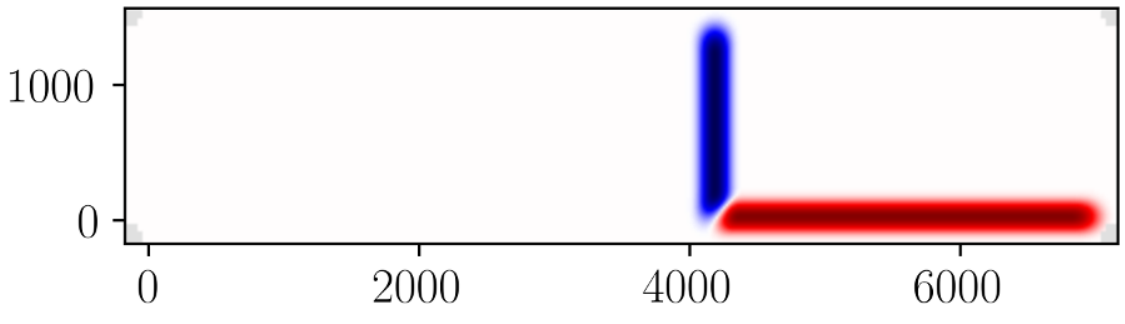}
\includegraphics[width=0.90\columnwidth]{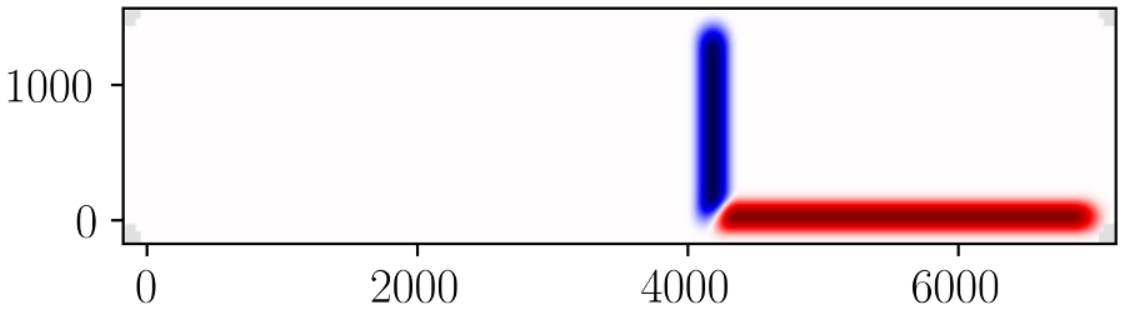}
\includegraphics[width=0.90\columnwidth]{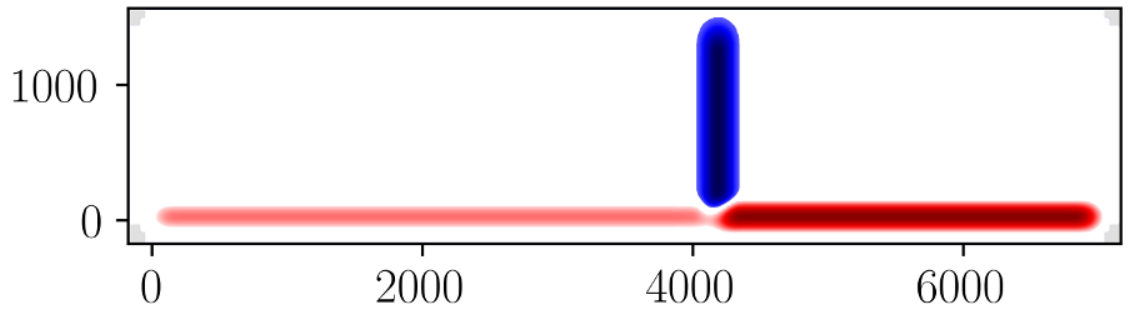}
  \caption{ Color map of the spin density in the full device at (top) $g_{\perp} = g_{\parallel} = 0$, 
  (middle) $g_{\parallel} = 0$, $g_{\perp} = 1.16$ meV and (bottom) $g_{\perp} = 0$, $g_{\parallel} = 0.35$ meV.
  The red (blue) colors indicate spin polarization along $s_z = \ua (s_{z} = \da)$. In all cases, spin up 
  electrons, injected from $S$, are incident upon the N-F-S junction. In the topological phase (top panel), 
  the electrons undergo Andreev reflection and spin down holes enter $D_{2}$. In the trivial phase (middle panel),
  the electrons undergo normal reflection and spin down electrons enter $D_{2}$. Finally, in the gapless phase
  (bottom panel), the electrons are partially reflected and partially transmitted through the superconductor, 
  entering both $D_{1}$ and $D_{2}$. Note that there is no net current between $D_{1}$ and $D_{2}$ at the top edge in any case. }
  \label{F-device}
\end{figure}

\subsubsection*{Magnetic Perturbations}

We further add magnetic perturbations in the topological superconductor through 
$g_{\parallel} \sigma_z$ (boost) and $g_{\perp} \sigma_x$ (mass). Note that these terms are also introduced throughout the
setup, but primarily affect only the gapless edge modes. While the effect of $g_{\perp}$ has been explored in detail~\cite{Alicea2011,Alicea2012},
the effect of $g_{\parallel}$ is not well studied. Fig.~S3 shows the evolution of the band structure 
of the 2D (top panels) and 1D (bottom panels) models with boost. The bottom panels show the evolution
of the edge mode dispersion and the transition into the regime of gapless superconductivity (as described in the main text). 
Clearly, the edge bands of the 2D model also evolve in the same fashion, again confirming that the 
magnetic perturbations affect only the helical edge modes.

\subsubsection*{Full Device}

Having verified that we can faithfully simulate topological superconductivity on the QSH edge (despite adding a pairing
term in the full QSH region), we turn to the simulation of the full device shown in Fig.~1 of the main text. We create 
a $700$ $\times 140$ square lattice and define the model described above on it. To model the region in the quantum anomalous Hall 
(QAH) phase, an additional exchange term $g_{0} \sigma_{z} \tau_{z}$ is added to the Hamiltonian. Since 
$g_{0}$ increases the bulk gap in one spin sector and decreases the gap in the other one, it drives a 
topological phase transition in one of the spin sectors at sufficiently large values, giving rise to the 
QAH phase~\cite{Xing2013}. For our simulations, we impose $g_{0} = -12$ meV in a region of 
width $280$ sites at the right boundary. This defines the QAH region of the device. We use 
$g_{0} = 0$ in the rest of the device (thereby defining the QSH phase). 

Next, we add the superconducting pairing term in the full QSH region, and the magnetic perturbations in the entire setup (as explained above). 
For our simulations, we use $\Delta = 0.3$ meV. Finally, the ferromagnetic 
region (of width 20 lattice sites) is introduced, between the QSH and QAH regions, through the Zeeman term $M_x \sigma_{x}$, 
which hybridizes the helical edge modes and opens a mass gap. 
The gapped region helps to localize the Majorana mode at the boundary of the topological superconductor, and sharpens 
the zero energy resonance observed in transport. Here, we use $M_x = 0.85$ meV which is sufficiently large for this 
purpose [Fig.~1(a)]. Quantum transport through the device is simulated between the three leads ($S$, $D_{1}$ and $D_{2}$), which are
also described by the device Hamiltonian.

Fig.~S4 shows the spin density in the full device, computed in the three phases : topological (top panel), 
trivial (middle panel) and gapless (bottom panel). Clearly the source ($S$) injects spin-polarized electrons only at the
bottom edge, due to the chiral modes of the QAH region which is taken to be spin up polarized. The injected electrons are incident upon
the N-F-S junction (here, F is the ferromagnetic barrier) and may undergo normal/Andreev reflection to enter $D_2$ or 
transmission across the superconductor to enter $D_1$. As expected we find that in the topological phase (top panel), 
the electrons undergo perfect Andreev reflection and enter $D_2$ as holes. The Andreev nature of the reflection is seen in 
the differential conductance at zero energy (shown in Fig.~3 of the main text). On the other hand, in the trivial phase
(middle panel), the electrons undergo perfect normal reflection and enter $D_2$ as electrons. Most interestingly, in the
gapless phase (bottom panel), the electrons may partially transmit through the superconductor entering $D_{1}$. There is no
current along the top edge connecting $D_{1}$ and $D_{2}$ since both are grounded. 

\subsubsection*{Robustness to Disorder}

We repeat our numerical analysis of the setup in a non-ideal situation in order to demonstrate the advantage of our
proposal over conventional setups. For this purpose, we introduce both time-reversal invariant and 
time-reversal breaking disorder to the system. The former is added through the term, 
\begin{align}
  H_{U} = \sum_{\vec{r}} U(\vec{r}\,) \sigma_{0} \tau_{0}
\end{align}
where, $U(\vec{r}\,)$ is a random onsite potential drawn from a uniform distribution in the range 
$[-5, 5]$ meV. The potential on different sites is uncorrelated. The non-magnetic impurity potential is not 
sufficient to introduce spin-flip scattering in the device and is relatively innocuous by itself. 
Therefore we also introduce a ferromagnetic disorder described by, 
\begin{align}
  H_{M} = \sum_{\vec{r}} M(\vec{r}\,) \sigma_{x} \tau_{0}
\end{align}
where, $M(\vec{r}\,)$ is a random onsite spin-mixing term drawn from a uniform distribution in the range 
$[-0.3, 0.3]$ meV. Again the potential on different sites is uncorrelated. 
The non-uniform onsite potential and ferromagnetic terms allow random rotation of the electronic spin in the device. 

A crucial aspect of our proposal is the chiral injection of electrons arising from the presence of the QAH region. 
To better understand the interplay of edge mode chirality and disorder, we also performed numerical simulations
on an alternate geometry, adapted from the setup in Ref.~\cite{Beenakker2013}, without the QAH region. The source injects electrons 
directly into the helical edge modes of the 
QSH phase. The presence of disorder-induced-backscattering reduces the width of the zero bias peak in this case~\cite{Beenakker2013}
and renders a clean observation of Majorana modes difficult. In contrast, our proposal is much more robust to 
the effects of random scattering.

\bibliographystyle{apsrev}